# The Hamiltonian brain: efficient probabilistic inference with excitatory-inhibitory neural circuit dynamics

Laurence Aitchison[1]* and Máté Lengyel[2,3]

[1] Gatsby Computational Neuroscience Unit, University College London, London, United Kingdom

[2] Computational & Biological Learning Lab, Department of Engineering, University of Cambridge, Cambridge, United Kingdom

[3] Department of Cognitive Science, Central European University, Budapest, Hungary

* laurence.aitchison@gmail.com

## Abstract

Probabilistic inference offers a principled framework for understanding both behaviour and cortical computation. However, two basic and ubiquitous properties of cortical responses seem difficult to reconcile with probabilistic inference: neural activity displays prominent oscillations in response to constant input, and large transient changes in response to stimulus onset. Indeed, cortical models of probabilistic inference have typically either concentrated on tuning curve or receptive field properties and remained agnostic as to the underlying circuit dynamics, or had simplistic dynamics that gave neither oscillations nor transients. Here we show that these dynamical behaviours may in fact be understood as hallmarks of the specific representation and algorithm that the cortex employs to perform probabilistic inference. We demonstrate that a particular family of probabilistic inference algorithms, Hamiltonian Monte Carlo (HMC), naturally maps onto the dynamics of excitatory-inhibitory neural networks. Specifically, we constructed a model of an excitatory-inhibitory circuit in primary visual cortex that performed HMC inference, and thus inherently gave rise to oscillations and transients. These oscillations were not mere epiphenomena but served an important functional role: speeding up inference by rapidly spanning a large volume of state space. Inference thus became an order of magnitude more efficient than in a non-oscillatory variant of the model. In addition, the network matched two specific properties of observed neural dynamics that would otherwise be dif-ficult to account for in the context of probabilistic inference. First, the frequency of oscillations as well as the magnitude of transients increased with the contrast of the image stimulus. Second, excitation and inhibition were balanced, and inhibition lagged excitation. These results suggest a new functional role for the separation of cortical populations into excitatory and inhibitory neurons, and for the neural oscillations that emerge in such excitatory-inhibitory networks: enhancing the efficiency of cortical computations.

## Author Summary

Our brain operates in the face of substantial uncertainty due to ambiguity in the inputs, and inherent unpredictability in the environment. Behavioural and neural evidence indicates that the brain often uses a close approximation of the optimal strategy, probabilistic inference, to interpret sensory inputs and make decisions under uncertainty. However, the circuit dynamics underlying such probabilistic computations are unknown. In particular, two fundamental properties of cortical responses, the presence of oscillations and transients, are difficult to reconcile with probabilistic inference. We show that excitatory-inhibitory neural networks are naturally suited to implement a particular inference algorithm, Hamiltonian Monte Carlo. Our network showed oscillations and transients like those found in the cortex and took advantage of these dynamical motifs to speed up inference by an order of magnitude. These results suggest a new functional role for the separation of cortical populations into excitatory and inhibitory neurons, and for the neural oscillations that emerge in such excitatory-inhibitory networks: enhancing the efficiency of cortical computations.

## Introduction

Uncertainty plagues neural computation. For instance, hearing the rustle of an animal at night, it may be impossible to ascertain the species, and thus whether or not it is dangerous. One approach in this scenario is to respond based on a point estimate, usually the single most probable explanation of our observations. However, this leads to a problem: if the probability of the animal being dangerous is below 50%, then the single most probable explanation is that the animal is harmless; and considering only this explanation, and thus failing to respond, could easily prove fatal. Instead, to respond appropriately, it is critical to take uncertainty into account by also considering the possibility of there being a dangerous animal, given the rustle and any other available clues.

The optimal way to perform computations and select actions under uncertainty is to represent a probability distribution that quantifies the probability with which each scenario may describe the actual state of the world, and update this probability distribution according to the laws of probability, i.e. by performing Bayesian inference. Human behaviour is consistent with Bayesian inference in many sensory [1, 2, 3, 4], motor [5, 6] and cognitive [7, 8, 9] tasks. There is also evidence that probabilistic inference is performed already in early sensory cortical areas [10, 11]. In particular, simple cells in the primary visual cortex (V1) respond maximally to Gabor filter-like stimuli (i.e. edges), which have been shown to provide the most parsimonious explanation of natural images in probabilistic theories of visual processing [12] (or mathematically equivalent regularisation-based approaches [13]). Furthermore, more complex probabilistic models can account for contrast invariant tuning [14] and complex cell properties [15], as well as surround-suppression effects in neural data and behaviour [16].

The apparent success of probabilistic inference in accounting for a diverse set of experimental observations raises the question of how neural systems might represent and compute with uncertainty [17]. Nevertheless, traditional models of neural computation ignore uncertainty, and instead rely on circuit dynamics that find the single best explanation for their inputs [18, 13, 19]. More recent approaches do allow for the representation of uncertainty, including distributional [20], doubly distributed [21], and probabilistic population codes [22, 23, 24], or sampling-based network dynamics [25, 26, 11]. However, none of these previous models capture the rich dynamics of cortical responses. In particular, neural activities in the cortex show prominent intrinsic oscillations [27], and large transient changes in response to stimulus onset, which are observed in V1 [28, 29, 30], and other cortical areas [31, 32]. In contrast, existing neural models of probabilistic inference either have no dynamics and so predict stationary responses to a fixed stimulus, or they have gradient ascent-like dynamics that display neither oscillations nor transients, and eventually also converge to a steady-state response for a fixed input. Moreover, these models typically violate Dale's law, by having neurons with both excitatory and inhibitory outputs. While there have been excitatory-inhibitory (EI) networks models that did capture some of these aspects of cortical dynamics, these have rarely been linked to any particular computation (but see [33, 34]), let alone probabilistic inference.

Here, we present an EI neural network model of V1 that performs probabilistic inference while retaining a computationally useful representation of uncertainty, and has rich, cortex-like dynamics, including oscillations and transients. In particular, our network uses a sampling-based representation of uncertainty [25, 35, 11], such that at any time it represents a single plausible interpretation of the input, and as time passes it sequentially samples many different interpretations. In other words, the network represents the probability of different scenarios implicitly, by the frequency with which it visits their representations via its dynamics. For instance, in the example above, neural activity at one moment would represent "dangerous", then "not dangerous" at some later time, and then "dangerous" again, such that a decision about how to behave can then be made based on the proportion of the time neural activity represents "dangerous" vs. "not dangerous". Thus, a fundamental consequence of a sampling-based representation for neural dynamics is that whenever there is uncertainty, neural activity will not settle down to a single fixed point but instead, it will continue to move between patterns representing the different possible states of the world. More specifically, an *efficient* sampling-based representation requires this continuous movement across state space to be such that the rate at which (statistically independent) samples are generated by the dynamics is as high as possible. We show that EI networks are ideally suited to achieve efficient sampling by implementing a powerful family of probabilistic inference algorithms, Hamiltonian Monte Carlo (HMC) [36, 37].

HMC is based on the idea that it is possible to sample from a probability distribution by setting up a dynamical system whose dynamics is Hamiltonian (Fig. 1A). The state of such a system behaves as a particle moving on a (high dimensional) surface, with momentum. The surface determines the potential energy of the particle, corresponding to the negative logarithm of the probability distribution that needs to be sampled (such that high probability states correspond to low potential energy). These dynamics speed up inference because the momentum of the system prevents the random walk behaviour plaguing many other sampling-based inference schemes. In particular, the particle will accelerate as it heads towards the minimum of the potential energy landscape, but once it reaches that point, it will have a large momentum, so it will keep moving out the other side (Fig. 1A-D). Our key insight is that HMC dynamics are naturally implemented by the interactions of recurrently coupled excitatory and inhibitory populations in cortical circuits. Due to these interactions, our network possessed inherently oscillatory dynamics. Crucially, these oscillations were ideal for speeding up inference, as they moved rapidly across the state space and hence represented a whole range of plausible interpretations efficiently.

In the following, we first define the statistical model of natural visual scenes that served as the testbed for our simulations of V1 dynamics. We then describe the HMC-based neural network that implemented sampling under this statistical model. We demonstrate that our dynamics sample more rapidly than noisy gradient ascent (also known as Langevin dynamics), and therefore that the presence of oscillations and transients in our network speeds up inference. Next, we show by both theoretical analysis and simulation that our sampler reproduces three properties of experimentally observed cortical dynamics. First, our sampler has balanced excitation and inhibition, with inhibition lagging excitation [38]. Second, our sampler oscillates, and the oscillation frequency increases with stimulus contrast [30, 39]. Third, there is a transient increase in firing rates upon stimulus onset, and the magnitude of this transient is also modulated by stimulus contrast [30]. Thus, our work provides a principled unifying account of these dynamical motifs by relating them to a fundamental class of cortical computations: probabilistic inference.

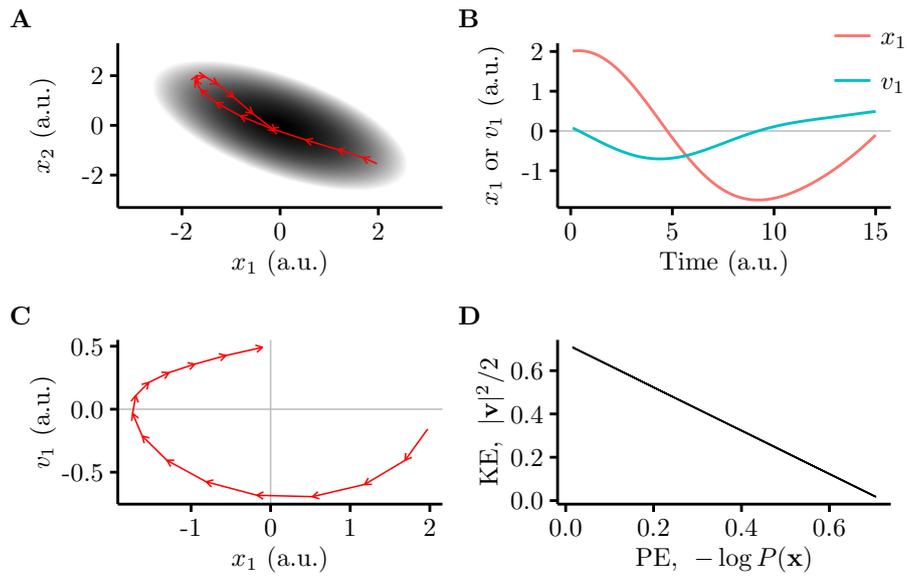

**Figure 1.** An example of Hamiltonian dynamics. **A.** Movement of a particle under Hamiltonian dynamics (i.e. with momentum) on a two-dimensional quadratic potential energy landscape (greyscale, darker means lower energy) corresponding to a multivariate Gaussian probability density. The red arrows show the trajectory, with each arrow representing an equal time interval. Note that the particle does not just go to the lowest potential energy location: it picks up momentum (kinetic energy) as it moves, leading it to oscillate around the energy well. **B.** A plot of position (red) and velocity (blue, the derivative of position) along one dimension. **C.** Plotting velocity and position directly against each other reveals explicitly that the dynamics of the system is similar to that of a harmonic oscillator. **D.** Plotting kinetic energy (KE) against potential energy (PE) reveals an exchange between kinetic energy and potential energy that contributes to the system's oscillatory behaviour.

Table 1. Values of the parameters used in our simulations.

| Parameter | Value | Role |
|---|---|---|
| $\mathbf{C}$ | $\left(1 - \sigma_{\mathbf{x}}^2\right)\left(\mathbf{A}^T\mathbf{A}\right)^{-1}$ | prior covariance of $\mathbf{u}$ |
| $\mathbf{A}$ | See Fig. 2B and Methods | edge-detecting filters represented by model neurons |
| $\sigma_{\mathbf{x}}^2$ | 0.1 | variance of observation noise |
| $\tau$ | 10 ms | membrane time constant |
| $\rho^2$ | 13 s$^{-1}$ | rate at which stochastic vesicle release injects noise |
| $\mathbf{W}_{\mathrm{uu}}, \mathbf{W}_{\mathrm{uv}}$, etc. | See Methods | recurrent connection weights in the network |

See Methods for details of the procedure used to determine the parameters. Oscillation frequency in the network was jointly determined by several of these parameters (see Eq. 8), the timescale of transients was mainly determined by $\rho$ (see S1 Figure).

# Results

## The Gaussian scale mixture model and V1 responses

In order to model the dynamics of V1 responses, we adopted a statistical model that has been widely used to capture the statistics of natural images and consequently to account for the *stationary* responses of V1 neurons in terms of probabilistic inference. We extended this model to account for the *dynamics* of V1 responses.

The Gaussian scale mixture (GSM) model is relatively simple, yet captures some fundamental higher-order statistical properties of natural image patches by introducing latent variables, $\mathbf{u}$, coordinating the linear superposition of simple edge features and an additional latent variable, $z$, determining the overall contrast level of the image patch [40] (Fig. 2A). Formally, the probabilistic generative model can be written as

$$P(\mathbf{u}) = \mathcal{N}(\mathbf{u}; \mathbf{0}, \mathbf{C}) \qquad (1)$$
$$P(z) = \mathcal{T}(z; 0, 1, 0) \qquad (2)$$
$$P(\mathbf{x}|\mathbf{u}, z) = \mathcal{N}(\mathbf{x}; z\mathbf{A}\mathbf{u}, \sigma_{\mathbf{x}}^2 \mathbf{I}) \qquad (3)$$

where $\mathcal{N}(\cdot; \boldsymbol{\mu}, \boldsymbol{\Sigma})$ is a multivariate distribution with mean $\boldsymbol{\mu}$ and covariance $\boldsymbol{\Sigma}$, $\mathcal{T}(\cdot; \mu, \sigma^2, \theta)$ is a truncated (univariate) normal distribution with mean $\mu$ and variance $\sigma^2$ truncated below threshold $\theta$ (so that, in our case, $z$ is non-negative), $\mathbf{x}$ is the grey levels of pixels in an image patch, the columns of $\mathbf{A}$ include the edge-like features whose combinations are used to explain images (Fig. 2B), $\mathbf{C}$ describes their prior covariance (which is fitted to whitened data), and $\sigma_{\mathbf{x}}^2 = 0.1$ is the level of noise present in the images. (See Table 1 for all parameters in the model, and Methods for details of the procedure used to set them.)

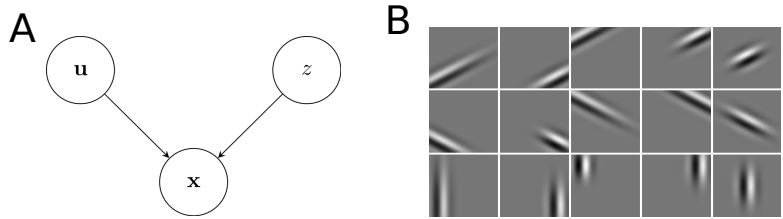

**Figure 2. A.** The graphical model representation of the Gaussian scale mixture model. The distribution over the observations (images), $\mathbf{x}$, depends on two latent variables, $z$ and $\mathbf{u}$. The vector $\mathbf{u}$ represents the intensity of edge-like features (see panel B) in the images. The positive scalar $z$ represents the overall contrast level in the image. **B.** The basis functions represented by $\mathbf{u}$ were 15 Gabor filters centred at five different locations, and with three different orientations.

Crucially, assuming that V1 simple cell activities represent values of $\mathbf{u}$ sampled from the posterior over $\mathbf{u}$ given an input $\mathbf{x}$ under the GSM, $P(\mathbf{u}|\mathbf{x})$, provides a natural account for a number of empirical observations. (Conversely, inference of $z$ may provide an account of complex-cell activations [41, 42, 43], which we did not study in further detail here.) In particular, the posterior mean of $\mathbf{u}$, represented by the mean of model neuron activities, matches the across-trial average responses of simple cells in V1 [14, 44]. Moreover, it can also be shown that the posterior variance of $\mathbf{u}$, represented by the variance of model neuron activities, captures important aspects of the across-trial variance of V1 responses [11], namely the quenching of neural variability with stimulus onset [45]. This is because, in the no-stimulus condition, we have a blank image, $\mathbf{x} = \mathbf{0}$. Under the GSM, $\mathbf{x} \approx z\mathbf{A}\mathbf{u}$, so while it is possible to explain a blank image by setting every single element of $\mathbf{u}$ very close to 0 (or, more generally, tuning $\mathbf{u}$ to be in the nullspace of $\mathbf{A}$), a far more parsimonious, and probable, explanation is that $z$ (a single scalar) is close to 0. Importantly, if $z$ is close to 0, then $\mathbf{x}$ does not constrain $\mathbf{u}$. Plausible values for $\mathbf{u}$ therefore cover a broad range (defined by the prior over $\mathbf{u}$), so $\mathbf{u}$ and hence neural activity, can be highly variable. In contrast, if there is a stimulus, $\mathbf{x} \not\approx \mathbf{0}$, we must also have $z \not\approx 0$, in which case $\mathbf{x}$ tightly constrains the range of plausible values of $\mathbf{u}$ (as $\mathbf{x} \approx z\mathbf{A}\mathbf{u}$), leading to lower variability. Moreover, the model naturally implements a form of divisive gain control: a very large $\mathbf{x}$ can be accounted for by making $z$, rather than $\mathbf{u}$, large [46]. This agreement between the probabilistic model and empirically observed patterns of neural activity is our key motivation for choosing the GSM model as our testbed and asking what plausible neural network dynamics may be appropriate for sampling from its posterior distribution.

## Hamiltonian Monte Carlo in an EI network

To ensure efficient sampling from the posterior, we constructed network dynamics based on the core principles of HMC sampling. The efficiency of HMC stems from its ability to speed up inference by preventing the random walk behaviour plaguing other sampling-based inference schemes. In particular, it introduces auxiliary variables to complement the 'principal' variables whose value needs to be inferred (**u** in the case of the GSM). Although this extension of the state space seemingly makes computations more challenging, it allows inference to be substantially more efficient when dynamical interactions between the two groups of variables are set up appropriately.

We noted that the particular interaction between principal and auxiliary variables required by HMC dynamics is naturally implemented by the recurrently connected excitatory and inhibitory populations of cortical circuits. Thus, the dynamics of our two-population neural network that sampled from the GSM posterior were (Fig. 3, see Methods for a full derivation):

$$\dot{\mathbf{u}} = \frac{1}{\tau} \left[ \mathbf{W}_{uu}\mathbf{u} - \mathbf{W}_{uv}\mathbf{v} + \tfrac{1}{2}\tau\rho^2 \mathbf{I}_{\text{input}} \right] + \rho\boldsymbol{\eta}_u \quad (4)$$

$$\dot{\mathbf{v}} = \frac{1}{\tau} \left[ \mathbf{W}_{vu}\mathbf{u} - \mathbf{W}_{vv}\mathbf{v} - \mathbf{I}_{\text{input}} \right] + \rho\boldsymbol{\eta}_v \quad (5)$$

where $\boldsymbol{\eta}_u$ and $\boldsymbol{\eta}_v$ denotes standard normal white noise (or, more precisely, the differential of a Wiener processes), the **W** matrices are the recurrent synaptic weight matrices between the two populations of cells (defined in the Methods), such that all their elements are positive, and

$$\mathbf{I}_{\text{input}} = \frac{z}{\sigma_x^2} \mathbf{A}^T \left( \mathbf{x} - z\mathbf{A}\mathbf{u} \right) - \mathbf{C}^{-1}\mathbf{u} \quad (6)$$

is an input current. Under these dynamics, the principal $u_i$ and auxiliary variables $v_i$ corresponded to the membrane potentials of individual neurons (or the average membrane potential of small populations of cells), and for any input **x**, the stationary distribution of **u** was guaranteed to be identical to the corresponding posterior distribution under the GSM.

Network dynamics consisted of three components. First, recurrent dynamics implementing HMC was specified by the first two terms in Eqs (4) and (5), $\mathbf{W}_{uu}\mathbf{u} - \mathbf{W}_{uv}\mathbf{v}$ and $\mathbf{W}_{vu}\mathbf{u} - \mathbf{W}_{vv}\mathbf{v}$. As the elements of the **W** matrices were all positive (see above), the recurrent circuit implied by these dynamics had an EI structure, with **u** corresponding to excitatory cells and **v** to inhibitory cells.

Second, there was an input current $\mathbf{I}_{\text{input}}$, whose strength was scaled by the (inferred) level of contrast, $z$ (Eq. 6). Note again that while this signal might increase with $z$, it is a prediction error, so it has a highly non-trivial relationship with the resulting response. In fact, it can be shown that the response actually saturates as contrast increases (and results in tuning curves with contrast invariant width) [11]. This input current specified the probabilistic model by conveying a prediction error, i.e. the difference between the input image, **x**, and the image predicted by the current activities of the excitatory neurons, $z\mathbf{A}\mathbf{u}$, plus a term penalizing the violation of prior expectations about **u**. While the key focus of our paper is the EI circuit implementing HMC, rather than the specific form for the input (of which the details depend on the underlying probabilistic model, here the admittedly simplified GSM model), we suggest a potential implementation of $\mathbf{I}_{\text{input}}$ by a separate population of neurons directly representing the prediction error $(\mathbf{x} - z\mathbf{A}\mathbf{u})$ as in theories of predictive coding [18]. Such cells (perhaps in the lateral geniculate nucleus, LGN) would have an excitatory connection from upstream areas (the retina), representing the data, and an inhibitory disynaptic connection from the excitatory cells, **u**. The output from these cells needs to excite the excitatory cells and inhibit the inhibitory cells of our circuit, which can again be implemented via disynaptic inhibition. This form of input is particularly well-suited to give strong, long-lasting activation of the EI circuit, as the increase in excitation reinforces the decrease in inhibition.

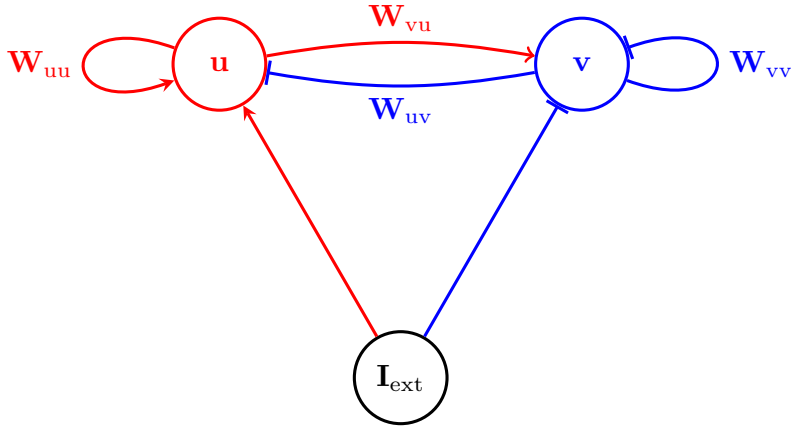

**Figure 3.** The architecture of the Hamiltonian network. The network consists of two populations of neurons, excitatory neurons with membrane potential **u**, and inhibitory neurons **v**, driven by external input $\mathbf{I}_{\text{input}}$. Neurons in the network are recurrently coupled by synaptic weights, $\mathbf{W}_{\text{uu}}$, $\mathbf{W}_{\text{uv}}$, $\mathbf{W}_{\text{vu}}$ and $\mathbf{W}_{\text{vv}}$. Red arrows represent excitation; blue bars represent inhibition.

Finally, the last term in Eqs (4) and (5) represented noise. Although these dynamics were clearly simplified in that they were fundamentally linear, such dynamical systems have been used to model a wide variety of neural processes [47, 48, 49]. Previous work has also shown that neurons combining firing-rate nonlinearities with short-term synaptic plasticity and dendritic nonlinearities can implement such effectively linear membrane potential dynamics [50, 51]. Moreover, such models have been found to provide a good match to the dynamics of cortical populations at the level of field potentials [52], calcium signals [53], and firing rate trajectories [54, 49]. We set the parameters of the network to lie in a biologically realistic regime (Table 1, Methods).

## Oscillations contribute to efficient sampling

When given an input image, our network exhibited oscillatory dynamics due to its intrinsic excitatory-inhibitory interactions (Fig. 4A). Intuitively, these oscillations were useful for inference as they allowed the network to cover a broad range of plausible interpretations of its input within each oscillation cycle. In order to assess more rigorously the computational use of these oscillations, we compared our network to a non-oscillatory counterpart, called Langevin sampling [55] (Methods). For a fair comparison of the two samplers, we set them up to sample from the same posterior, and we kept the noise level $\rho$ the same in them.

The Langevin sampler was constructed by setting the recurrent weights in our network ($\mathbf{W}$ matrices) to zero. Although, in general, a Langevin sampler can still have recurrent connectivity, at least among the principal cells (by interpreting the dependence of $\mathbf{I}_{\text{input}}$ on $\mathbf{u}$ as recurrent connections [56]), these recurrent connections are necessarily symmetric and therefore fundamentally different in nature from the EI interactions that we consider here. As a consequence, Langevin dynamics showed prominent random walk-like behaviour without oscillations (Fig. 4B). Comparing the autocorrelation functions for the Hamiltonian and Langevin samplers revealed that while their autocorrelation functions decayed at similar rates (controlled by the timescale of the stochastic, Langevin component), HMC had an additional, oscillating component, (Fig. 4C).

The oscillatory behaviour of our HMC sampler allowed it to explore a larger volume of state space in a fixed time interval than Langevin sampling (Fig. 4D-E). To compare the sampling performance of HMC and Langevin dynamics rigorously, we measured for both of them the error between a sample-based estimate of the posterior mean and the true mean of the posterior. The samples from the Hamiltonian sampler took very little time to give a good estimate of the mean (73 ms to get the mean square error to the level obtainable by a single statistically fair sample), whereas samples from the Langevin model took ∼4 times longer (273 ms, Fig. 4F). This difference indicated that our HMC-inspired sampler used limited noise far more efficiently than Langevin dynamics.

The efficiency of HMC is typically attributed to the suppression of the random walk behaviour of Langevin dynamics [37]. In our network, we were able to relate this effect more specifically to the appearance of oscillations. HMC dynamics had both an oscillatory and a stochastic component (Fig. 4A, C red), whereas Langevin dynamics had only the stochastic component, so that it performed simple noisy gradient ascent, without apparent oscillations (Fig. 4B, C blue). In particular, oscillations in the HMC sampler had a time scale that was a factor of 15 faster than that of the stochastic component shared with Langevin dynamics. This fast time constant of the HMC sampler, $\tau$, governed the effects of recurrent EI interactions, which were mediated by the $\mathbf{W}$ matrices that the Langevin sampler lacked (Eq. 32). These architectural and dynamical differences implied a fundamentally different strategy for exploring the state space of these networks. The fast oscillations in the HMC sampler deterministically explored states in $(\mathbf{u}, \mathbf{v})$-space that lay on an equiprobability manifold, while the slow time scale implied by the input noise served to change this manifold stochastically (Fig. 4D). Indeed, the autocorrelogram of the energy (log posterior probability) in the HMC sampler (Fig. 4C, red dashed curve) was identical to the Langevin envelope of the autocorrelogram of states (Fig. 4C, red solid curve), indicating that energy only changed on the slow time scale governed by this stochastic component and not on the fast time scale of oscillations. (Note that while moving along equiprobability contours in the full joint $(\mathbf{u}, \mathbf{v})$ space, HMC dynamics may still cross probability contours when projected to a low dimensional marginal, as shown in Fig. 4D.) In contrast, Langevin dynamics could only rely on this slow stochastic component resulting in slow movement across energy levels (Fig. 4C, blue dashed curve) and the state space (Fig. 4C, blue solid curve).

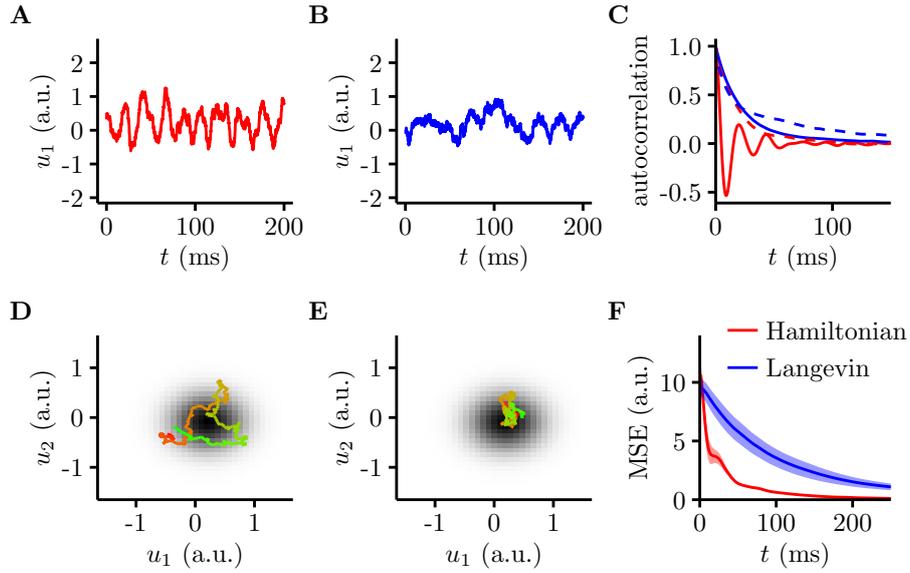

**Figure 4.** The Hamiltonian sampler is more efficient than a Langevin sampler. **A, B.** Example membrane potential traces for a randomly selected neuron in the Hamiltonian network (**A**) and the Langevin network (**B**). **C.** Solid lines: the autocorrelation of membrane potential traces in **A** and **B**, for Hamiltonian (red) and Langevin samplers (blue). Dashed lines: the autocorrelation of the joint (log) probability for Hamiltonian (red) and Langevin samplers (blue). Note that for the Hamiltonian sampler, the joint probability is over both **u** and **v**. **D, E.** Joint membrane potential traces from two randomly selected neurons in the Hamiltonian network (**D**) and the Langevin network (**E**), colour indicates time (from red to green, spanning 25 ms), grey scale map shows the (logarithm of the) underlying posterior (its marginal over the two dimensions shown). **F.** Normalised mean square error (MSE) between the true mean and the mean estimate from samples taken over a time $t$ for the Langevin (blue) and Hamiltonian dynamics (red), with 100 repetitions (mean $\pm$ 2 s.e.m.).

## Balance between excitation and inhibition

As we saw above, the advantage of HMC over Langevin dynamics could be attributed to the contribution of the recurrent connections, i.e. the $\mathbf{W}_{uu}\mathbf{u} - \mathbf{W}_{uv}\mathbf{v}$ and $\mathbf{W}_{vu}\mathbf{u} - \mathbf{W}_{vv}\mathbf{v}$ terms in the dynamics (Eq. 4 and 5), which respectively expressed the difference between net excitation and inhibition received by excitatory and inhibitory neurons. (Note that this difference was not affected by $\mathbf{I}_{\text{input}}$ as the prediction error conveyed by the input is zero on average for any input, by definition.) Importantly, for HMC to sample from the correct posterior, the dynamics of excitatory cells needed to track the prediction error conveyed by $\mathbf{I}_{\text{input}}$, for which the recurrent term needed to be zero on average, which in turn suggests that excitation and inhibition needed to track each other across different stimuli (Fig. 5A). Indeed, the only way we could obtain Hamiltonian dynamics that complied with Dale's law was if the activity of inhibitory cells tracked that of excitatory cells, i.e. if the network was balanced. As Langevin is equivalent to having these terms set to zero, for HMC to realize its advantage over Langevin, the variance of the recurrent term needed to be sufficiently large, which implied that the magnitudes of net excitation and net inhibition each needed to be large and momentarily imbalanced (Fig. 5B). These features, large excitatory and inhibitory currents that are tracking each other with momentary perturbations, are thought to be fundamental properties of the dynamical regime in which the cortex operates [38], and thus arise naturally from HMC dynamics in our EI network. Furthermore, as expected in a network with an EI architecture, excitation led inhibition in our network (Fig. 5C).

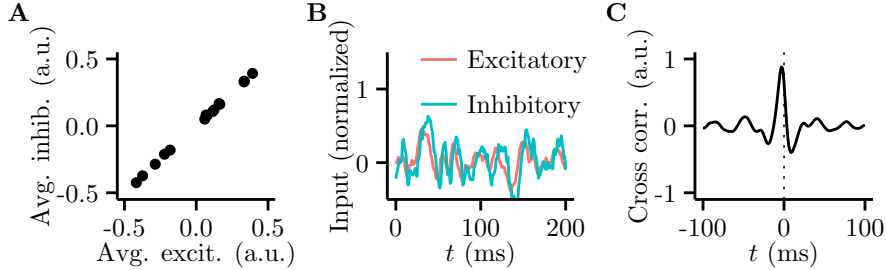

**Figure 5.** Excitation and inhibition are balanced in the Hamiltonian network. **A.** Trial-average excitatory input vs. trial-average inhibitory input across trials (dots) for a randomly selected individual cell in the network. **B.** Total inhibitory input to a single cell (blue) closely tracks but slightly lags total excitatory input (red) over the course of a trial. **C.** The cross-correlation between the average excitatory and average inhibitory membrane potentials shows a peak that is offset from 0 time.

### Stimulus-dependent oscillations

Oscillations are a ubiquitous property of cortical dynamics [57], and we have shown above that efficient sampling in HMC necessarily leads to oscillatory dynamics in general (Figs. 4-5). However, when applied specifically to perform inference based on visual images (Fig. 2), our model also reproduced some more specific and robust properties of gamma-band oscillations in V1, namely that the precise frequency of these oscillations increases with stimulus contrast [30, 39] (Fig. 6).

In order to extract an LFP from our model, in line with previous approaches (e.g. [58]), we computed the sum of membrane potentials of all cells. (Using the sum of input currents instead would have yielded qualitatively similar results.) The fact that LFP oscillations in our model were in the gamma band, i.e. around 40 Hz, was simply due to our choice of a realistic single neuron time constant, $\tau = 10$ ms. However, within this band, the modulation of the oscillation frequency by the contrast of the input image was a more specific characteristic of the dynamics of our network. As contrast increased, the amount of evidence to pin down **u** increased, and so the GSM posterior from which the dynamics needed to sample became tighter [11]. At the same time, the recurrent EI interactions of the HMC dynamics which gave rise to oscillations had a fixed time scale independent of the input (Eqs. 4 and 5). Using the same speed to traverse an equiprobability manifold of an increasingly tight posterior thus naturally led to increasing oscillation frequencies.

To further quantify this intuition, we simplified the dynamics of our network by incorporating the effects of inhibition directly into the equations describing the

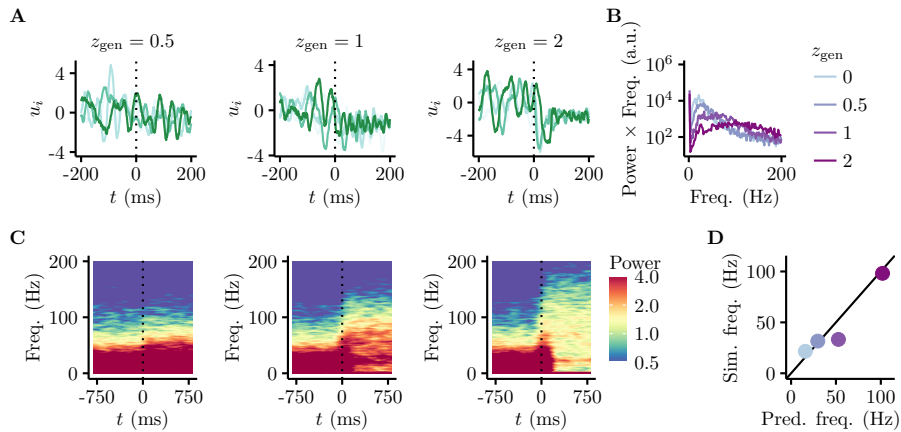

**Figure 6.** Oscillation frequency depends on stimulus contrast. **A.** The membrane potential response of one neuron to stimulus onset across 4 trials (coloured curves) shows that the variability decreases and the frequency increases as stimulus contrast increases. The true contrast of the underlying image increases left to right ($z_{\text{gen}} = 0.5$, 1, and 2). **B.** Power spectrum of the LFP (average membrane potentials) at different contrasts (coloured lines), showing that dominant oscillation frequency increases with contrast. Note that we plot power × frequency on the y-axis, in order to account for the fact that noise from a "scale-free" process has 1/f frequency dependence [59]. **C.** Time-dependent spectrum (Gaussian window, width 100 ms) of the LFP (contrast levels as in **A**). **D.** The simplified dynamics (x-axis, Eq. 8) accurately predicted the dependence of oscillation frequencies on contrast (colour code as in B) in the full network (y-axis).

dynamics of the excitatory cells (see Methods):

$$\ddot{\mathbf{u}} = -\frac{1}{\tau^2}\left(\frac{z^2}{\sigma_x^2} - \frac{1}{1-\sigma_x^2}\right)(\mathbf{u} - \bar{\mathbf{u}}) \tag{7}$$

where $\bar{\mathbf{u}} = \mathrm{E}\left[\mathbf{u}|\mathbf{x}, z\right]$ is the (stimulus-dependent) mean of the posterior over $\mathbf{u}$. This form explicitly exposes that our sampler (in the limit studied here) underwent regular harmonic oscillations, whose frequency increased with stimulus contrast, $z_{\text{gen}}$ (assuming that the inferred value of $z$ was sufficiently close to the actual stimulus contrast, i.e. $z \simeq z_{\text{gen}}$), as

$$f(z) = \frac{1}{2\pi\tau}\sqrt{\frac{z_{\text{gen}}^2}{\sigma_x^2} - \frac{1}{1-\sigma_x^2}} \tag{8}$$

Indeed, as predicted by these arguments, the network exhibited contrast-dependent oscillation frequencies both in its membrane potentials (Fig. 6A) and LFPs (Fig. 6B-C; note that in B, we account for the fact that a "scale-free" noise process has $1/f$ frequency dependence [59] by plotting power $\times$ frequency on the y-axis). Furthermore, the quantitative predictions made by Eq. 8 were in close agreement with the results of numerical simulations in the the full model, where $z$ is not fixed, but is inferred simultaneously with $\mathbf{u}$ (Fig. 6D).

## Stimulus-dependent transients

When we computed firing rates in the model by applying a threshold to membrane potentials (Eq. 60), our simulations showed large, contrast-dependent transient increases in population firing rate at stimulus onset (Fig. 7A). (Were we to consider the average membrane potential, this would not display such a large transient, because some neurons undergo positive transients, and others undergo negative transients, which cancel overall.) Such transients are also a widely observed characteristic of responses in V1 [29, 30] (as well as other sensory cortices [60, 32]). These transients were also inherent to the dynamics of our network and were not trivially predicted by simpler variants. For example, Langevin sampling did not give rise to any transient increase in firing rates — rates simply rose or fell towards their new steady state (Fig. 7B, most obvious for $z_{\text{gen}} = 0.5$). Even Hamiltonian dynamics did not necessarily yield transients. In particular, the full dynamics of our network inferred contrast, $z$, online together with the basis function intensities $\mathbf{u}$. Assuming instead that the brain knows $z = z_{\text{gen}}$, or uses a fixed value of $z$ sampled from $P(z|\mathbf{x})$, the dynamics became simple noisy harmonic motion. Although harmonic motion can lead to transients when initialised properly, the transients yielded by these dynamics were much smaller in magnitude which were near-impossible to detect in simulated population firing rates (Fig. 7C).

In order to understand how transients emerged in the full Hamiltonian dynamics of our network, sampling $\mathbf{u}$ and $z$ jointly, we focussed on the interaction between the dynamics of **u** and the inferred value of $z$. For analyzing the asymptotic behaviour in the previous section, we assumed that $z$ was constant (and equal to $z_{\text{gen}}$). However, in general, $z$ depended on the network's currently inferred value of **u**. In particular, $z$ and **u** jointly accounted for the total contrast content of the input image **x** (Eq. 3), and thus there was an inverse scaling between their magnitudes. Using the 1D variant of Eq. 7, $x \approx zAu$, so $z \approx x/Au$ (Fig. 7D). Here, we make use of a separation of time scales between the dynamics of $z$ and **u**, specifically that $z$ will attain its stationary value (distribution) much faster than **u**. This is because while the basis functions of $u_i$'s are localised Gabor filters, $z$ depends on the whole image patch (or, conversely, on all the $u_i$'s), which means that the sensory evidence for $z$ is much stronger than for **u**, and consequently its distribution is much narrower, giving strong prediction error signals which rapidly drive it to equilibrium. As $z$ effectively set the stiffness of the 'spring' underlying harmonic motions in our dynamics (Eq. 7), the system had high (restoring) acceleration for low values of $|u|$ and low accelerations for high values of $|u|$, resulting in high magnitude excursions in $u$ (Fig. 7E). Therefore, just after stimulus onset, $u$ was small, so there was a large force in the positive direction (due to the large stiffness), causing a large acceleration. Eventually, $u$ exceeded $\bar{u}$, but by that point the stiffness, and hence the restoring force had fallen, so the system's momentum allowed it to move a long distance, certainly further than if the spring constant had been fixed. This asymmetry in preferring upward to downward changes in $|u|$ was only relevant during initial transients as asymptotically the evidence in the image was sufficient to determine $z$ with high precision and so the dynamics of $u$ became approximately linear (as in Eq. 7). Thus, the timescale of the transient was determined by the timescale at which inferences about $z$ attained their stationary distribution, which in turn scaled with $\rho$ (S1 Figure).

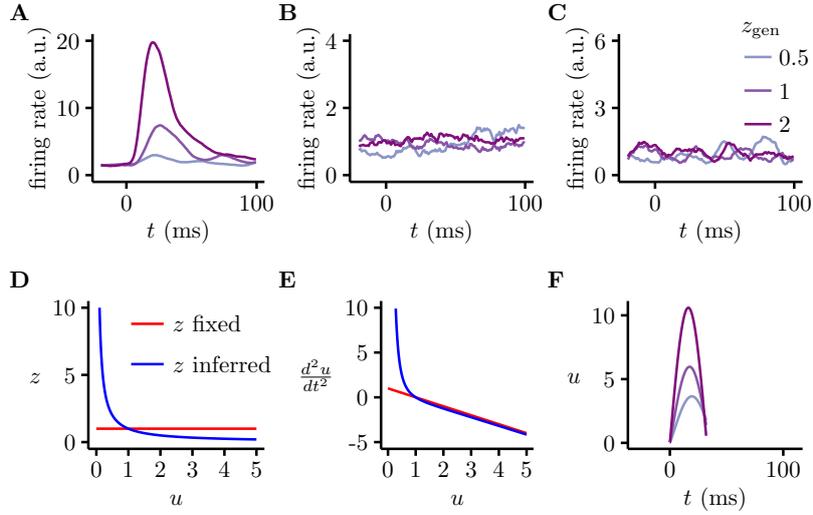

**Figure 7.** Large, contrast-dependent firing rate transients in the model. **A-C.** Transients (or lack thereof) at different contrast levels (colour) under the full dynamics (**A**), using Langevin dynamics (**B**), and under the full dynamics when the value of $z$ is fixed, $z = z_{\text{gen}}$ (**C**). Note different scales for firing rates in the three panels to better show the full range of firing rate fluctuations in each case. **D.** Dependence of the inferred value of contrast, $z$, on the currently inferred magnitude of basis function intensities, $u$, under the simplified dynamics (blue). For reference, red shows the value of $z$ when set to be fixed at $z = z_{\text{gen}}$. **E.** There is asymmetry in $\ddot{u}$ as a function of $u$, around the value of $u = \bar{u} = 1$, in the simplified model when $z$ is inferred (blue) but not when it is fixed (red). **F.** Transients predicted by the simplified dynamics (Eq. 9, with parameters as in Fig. 6D, and initial conditions $u(0) = 0.1$ and $\dot{u}(0) = 0$) are similar to transients under the full dynamics.

More formally, taking the 1D version of the simplified dynamics (Eq. 7), and substituting $z \approx x/Au$ gives

$$\ddot{u} = -\frac{1}{\tau^2} \left( \frac{x^2}{\sigma_{\text{x}}^2 A^2 u^2} + \frac{1}{1 - \sigma_{\text{x}}^2} \right) (u - \bar{u}) \tag{9}$$

Simulating this simplified dynamical system did indeed yield large transients (Fig. 7F) which matched full simulations (Fig. 7A) and recordings in macaque V1 [30] both in terms of the transient timescale (∼30 ms) and the dependence of transient magnitude on contrast level (values of $z_{\text{gen}}$). The fact that these large transients were retained in the model after such severe approximations indicated that they were robust to the exact method used for determining $z$, as long as it ensured that $z$ was consistent with both **x** and **u**.

# Discussion

Previously proposed mechanisms by which the cortex could either represent and manipulate uncertainty or just find the most probable explanation for sensory data failed to explain the richness of cortical dynamics. In particular, these models either had no dynamics or only gradient ascent-like dynamics, in contrast to neural activity in the cortex that displays oscillations in response to a fixed stimulus, and large transients in response to stimulus onset. Moreover, these models typically violated Dale's law, by having neurons whose outputs were both excitatory and inhibitory. We demonstrated that it was, in fact, possible to perform probabilistic inference in an EI network that displayed oscillations and transients. Moreover, having oscillations actually improved the network, in that it was able to perform inference faster than networks that did not have oscillations. Our model displayed four further dynamical properties that did not appear, at first, to be compatible with probabilistic inference: excitation and inhibition were balanced at the level of individual cells [38], inhibition lagged excitation [38], oscillation frequency increased with stimulus contrast [30], and there were large transients upon stimulus onset which also scaled with contrast [28, 29, 30]. In sum, we have given an approach by which successful, inference-based models of stationary activity distributions in V1 (e.g. [11]) can be extended to match the dynamics of neural activity.

Our work suggests a new functional role for cortical oscillations, and for inhibitory neurons that are involved in their generation: speeding up inference. We have demonstrated this role in the specific context of V1, but our formalism is readily applicable to other cortical areas in which probabilistic inference is supposed to take place, and similar stimulus-controlled transients and oscillations can be observed [61, 62]. Neural oscillations and probabilistic inference have been linked previously, albeit in the hippocampus rather than sensory cortices [63]. The main differences between the two approaches are that in previous work, oscillations were controlled entirely externally, and implemented (approximately) an augmented sampling scheme known as tempered transitions [64], whereas our work builds on the theory of Hamiltonian Monte Carlo [37] to construct network dynamics that are intrinsically oscillating. This allowed us to study the effects of the stimulus on these oscillations that previous approaches could not address. Computationally, Hamiltonian Monte Carlo and annealing-based techniques, such as tempered transitions, have complementary advantages in allowing network dynamics to respectively explore a given posterior mode or traverse different modes efficiently. Thus, a combination of these different approaches may account for concurrent cortical oscillations at different frequencies.

While the statistical model of images underlying our network was able to capture some interesting properties of the statistics of natural images, it was nevertheless clearly simplified, in that e.g. it did not capture any notion of objects, or occlusion. Once such higher-order features are incorporated into the model, we expect a variety of interesting new dynamical properties to emerge. For example, there should be strong statistical relationships between low-level variables describing a single object, and hence strong dynamical relationships, including synchronisation, between neurons representing different parts of the same object [65, 66]. In the extreme, we might expect to see coherent oscillations between neurons representing the same object, providing a principled unifying perspective of bottom-up (e.g. contrast) and top-down influences (e.g. "binding by synchrony") on cortical oscillations [67].

It will also be important to understand how local learning rules, modelling synaptic plasticity, may be able to set up the weight matrices that we found were necessary for implementing efficient Hamiltonian dynamics. For example, there might be two sets of learning rules operating in parallel, one set of rules which learns that statistical structure of the input, perhaps mainly through the plasticity of excitatory-to-excitatory connections [68], and another which tunes network dynamics, perhaps primarily by inhibitory plasticity mechanisms, to speed up the inference process, without altering the sampled distribution [69].

Finally, while the type of linear membrane potential dynamics we used in our network could be implemented using firing rate non-linearities in combination with synaptic and dendritic nonlinearities [50, 51], it will nevertheless be important to understand whether it is possible to perform inference in networks with more realistic non-linearities.

## Methods

### Sampler derivation

The sampler was derived by combining an HMC step, and a Langevin step to add noise and ensure ergodicity. The most general equations describing HMC are given by

$$\dot{\mathbf{u}} = \frac{1}{\tau} \frac{\partial \log P(\mathbf{u}, \mathbf{v}|\mathbf{x}, z)}{\partial \mathbf{v}} \tag{10}$$

$$\dot{\mathbf{v}} = -\frac{1}{\tau} \frac{\partial \log P(\mathbf{u}, \mathbf{v}|\mathbf{x}, z)}{\partial \mathbf{u}} \tag{11}$$

For the HMC step, there is freedom to specify the distribution of the auxiliary variable, $P(\mathbf{v}|\mathbf{u}, \mathbf{x})$, and freedom to set the noise distribution. Typically, the distribution of the auxilliary variable is set to have $\mathbf{0}$ mean and be totally independent of $\mathbf{u}$, so that $P(\mathbf{v}|\mathbf{u}, \mathbf{x}, z) = P(\mathbf{v}) = \mathcal{N}(\mathbf{v}; \mathbf{0}, \mathbf{M}^{-1})$. However, we know that inhibitory cells do, in fact, respond to input. We therefore chose to use

$$P(\mathbf{v}|\mathbf{u}, \mathbf{x}, z) = P(\mathbf{v}|\mathbf{u}) = \mathcal{N}(\mathbf{v}; \mathbf{Bu}, \mathbf{M}^{-1}) \tag{12}$$

with a free choice for $\mathbf{B}$ and $\mathbf{M}$, which we will discuss below (Setting the parameters). This allowed us to split up these probability distributions into terms that are dependent, and independent, of the data, $\mathbf{x}$:

$$\dot{\mathbf{u}} = \frac{1}{\tau} \frac{\partial \log P(\mathbf{v}|\mathbf{u})}{\partial \mathbf{v}} \tag{13}$$

$$\dot{\mathbf{v}} = -\frac{1}{\tau} \frac{\partial \log P(\mathbf{v}|\mathbf{u})}{\partial \mathbf{u}} - \frac{1}{\tau} \frac{\partial \log P(\mathbf{u}|\mathbf{x},z)}{\partial \mathbf{u}} \tag{14}$$

In order to add noise without perturbing the stationary distribution, we perform a Langevin step, that is, we simultaneously add noise and take a step along the gradient of the log-probability. Notably, this introduces a new time constant $\tau_L$, that simply controls the rate at which noise is injected into the system. As such, $\tau_L$ is directly related to $\rho$,

$$\rho = \sqrt{\frac{2}{\tau_L}} \tag{15}$$

The dynamics therefore become

$$\dot{\mathbf{u}} = \frac{1}{\tau} \frac{\partial \log P(\mathbf{v}|\mathbf{u})}{\partial \mathbf{v}} + \frac{1}{\tau_L} \frac{\partial \log P(\mathbf{u},\mathbf{v}|\mathbf{x},z)}{\partial \mathbf{u}} + \sqrt{\frac{2}{\tau_L}}\boldsymbol{\eta}_\mathrm{u} \tag{16}$$

$$\dot{\mathbf{v}} = -\frac{1}{\tau} \frac{\partial \log P(\mathbf{v}|\mathbf{u})}{\partial \mathbf{u}} - \frac{1}{\tau} \frac{\partial \log P(\mathbf{u}|\mathbf{x},z)}{\partial \mathbf{u}} + \frac{1}{\tau_L} \frac{\partial \log P(\mathbf{u},\mathbf{v}|\mathbf{x},z)}{\partial \mathbf{v}} + \sqrt{\frac{2}{\tau_L}}\boldsymbol{\eta}_\mathrm{v} \tag{17}$$

Again, we can break up the $P(\mathbf{u},\mathbf{v}|\mathbf{x},z)$ terms into terms that are dependent, and independent, of $\mathbf{v}$:

$$\dot{\mathbf{u}} = \frac{1}{\tau} \frac{\partial \log P(\mathbf{v}|\mathbf{u})}{\partial \mathbf{v}} + \frac{1}{\tau_L} \frac{\partial \log P(\mathbf{v}|\mathbf{u})}{\partial \mathbf{u}} + \frac{1}{\tau_L} \frac{\partial \log P(\mathbf{u}|\mathbf{x},z)}{\partial \mathbf{u}} + \sqrt{\frac{2}{\tau_L}}\boldsymbol{\eta}_\mathrm{u} \tag{18}$$

$$\dot{\mathbf{v}} = -\frac{1}{\tau} \frac{\partial \log P(\mathbf{v}|\mathbf{u})}{\partial \mathbf{u}} + \frac{1}{\tau_L} \frac{\partial \log P(\mathbf{v}|\mathbf{u})}{\partial \mathbf{v}} - \frac{1}{\tau} \frac{\partial \log P(\mathbf{u}|\mathbf{x},z)}{\partial \mathbf{u}} + \sqrt{\frac{2}{\tau_L}}\boldsymbol{\eta}_\mathrm{v} \tag{19}$$

Now, we compute these gradients, and convert them into a neural-network (see S1 Code)

$$\frac{\partial \log P(\mathbf{v}|\mathbf{u})}{\partial \mathbf{u}} = -\mathbf{M}(\mathbf{B}\mathbf{u} - \mathbf{v}) \tag{20}$$

$$\frac{\partial \log P(\mathbf{v}|\mathbf{u})}{\partial \mathbf{v}} = \mathbf{B}^T \mathbf{M}(\mathbf{B}\mathbf{u} - \mathbf{v}) \tag{21}$$

where the gradient of the posterior is the external input

$$\mathbf{I}_\mathrm{input} = \frac{\partial \log P(\mathbf{u}|\mathbf{x},z)}{\partial \mathbf{u}} = \frac{1}{\sigma_\mathrm{x}^2} z \mathbf{A}^T(\mathbf{x} - z\mathbf{A}\mathbf{u}) - \mathbf{C}\mathbf{u} \tag{22}$$

We can thus write the dynamics of our neural network as

$$\dot{\mathbf{u}} = \frac{1}{\tau}\left(\mathbf{W}_\mathrm{uu}\mathbf{u} - \mathbf{W}_\mathrm{uv}\mathbf{v} + \frac{\tau}{\tau_L}\mathbf{I}_\mathrm{input}\right) + \sqrt{\frac{2}{\tau_L}}\boldsymbol{\eta}_\mathrm{u} \tag{23}$$

$$\dot{\mathbf{v}} = \frac{1}{\tau}\left(\mathbf{W}_\mathrm{vu}\mathbf{u} - \mathbf{W}_\mathrm{vv}\mathbf{v} - \mathbf{I}_\mathrm{input}\right) + \sqrt{\frac{2}{\tau_L}}\boldsymbol{\eta}_\mathrm{v} \tag{24}$$

where

$$\mathbf{W}_{\mathrm{uu}} = \mathbf{B}^T\mathbf{M}\mathbf{B} - \frac{\tau}{\tau_L}\mathbf{M}\mathbf{B} \tag{25}$$

$$\mathbf{W}_{\mathrm{uv}} = \mathbf{B}^T\mathbf{M} - \frac{\tau}{\tau_L}\mathbf{M} \tag{26}$$

$$\mathbf{W}_{\mathrm{vu}} = \mathbf{M}\mathbf{B} + \frac{\tau}{\tau_L}\mathbf{B}^T\mathbf{M}\mathbf{B} \tag{27}$$

$$\mathbf{W}_{\mathrm{vv}} = \mathbf{M} + \frac{\tau}{\tau_L}\mathbf{B}^T\mathbf{M} \tag{28}$$

Finally, we substitute $\tau_L = 2/\rho^2$.

### Sampling $z$

The brain does not know $z_{\text{gen}}$, so it must infer $z$ together with $\mathbf{u}$. We therefore inferred $z$ and $\mathbf{u}$ in parallel, using an additional HMC sampler for $z$.

In particular, we simply extended the dynamics with an additional element for $z$:

$$\dot{z} = \frac{1}{\tau}\left(W_{zz}z - W_{zv}v + \frac{\tau}{\tau_L}I_{\text{input}}\right) + \sqrt{\frac{2}{\tau_L}}\eta_z \tag{29}$$

$$\dot{v} = \frac{1}{\tau}\left(W_{vz}z - W_{vv}v - I_{\text{input}}\right) + \sqrt{\frac{2}{\tau_L}}\eta_v \tag{30}$$

where $W$ is defined as above, with $B = M = 1$, and

$$I_{\text{intput}} = \frac{\partial \log P(\mathbf{u}, z, \mathbf{x})}{\partial z} = \frac{1}{\sigma_{\mathrm{x}}^2}(\mathbf{A}\mathbf{u})^T(\mathbf{x} - z\mathbf{A}\mathbf{u}) - z \tag{31}$$

### Langevin sampler

By setting the weight matrices implementing HMC, $\mathbf{W}$, to $\mathbf{0}$, we obtain the Langevin step:

$$\dot{\mathbf{u}} = \frac{1}{\tau_L}\mathbf{I}_{\text{input}} + \sqrt{\frac{2}{\tau_L}}\boldsymbol{\eta}_{\mathrm{u}} \tag{32}$$

### Setting the parameters

The GSM model has three parameters, the Gabor features, $\mathbf{A}$, the covariance matrix, $\mathbf{C}$, and the observation noise, $\sigma_{\mathrm{x}}^2$. We set $\mathbf{A}$ using known properties of the visual system: the Gabor filters-like receptive fields of V1 simple cells. In particular, we define $\mathbf{A}$ as a bank of Gabor filters at three orientations ($0$, $\pi/3$ and $2\pi/3$), five locations (the centre, and corners, $1/6$ image-widths from the edge, where all measurements are in units of image height = image width). The Gaussian envelope of the Gabors had minor axis $0.1$, and major axis uniformly distributed from $0.1$ to $0.5$ (where these measurements are in units of image width, and give the standard deviation along the relevant axis), and the sinusoid had wavelength $0.13$ image-widths.

We can set $\mathbf{C}$ using the value for $\mathbf{A}$, and the fact that retina and LGN are known to whiten visual input [70]. For a particular image, $\mathbf{x}$, and inferred contrast level, $z$, the posterior is

$$P(\mathbf{u}|\mathbf{x}, z) = \mathcal{N}\left(\mathbf{u}; \tfrac{z}{\sigma_\mathrm{x}^2}\boldsymbol{\Sigma}(z)\mathbf{A}^T\mathbf{x}, \boldsymbol{\Sigma}(z)\right) \tag{33}$$

where

$$\boldsymbol{\Sigma}(z) = \left(\mathbf{C}^{-1} + \tfrac{z^2}{\sigma_\mathrm{x}^2}\mathbf{A}^T\mathbf{A}\right)^{-1} \tag{34}$$

We know that the average posterior equals the prior [71, 10], and so the prior covariance $\mathbf{C}$ should match the average posterior covariance (averaging over data, $\mathbf{x}$, and other latent variables, $z$), i.e.

$$\mathbf{C} = \mathrm{E}\left[\mathbf{u}\mathbf{u}^T\right] = \mathrm{E}\left[\tfrac{z^2}{\sigma_\mathrm{x}^4}\boldsymbol{\Sigma}(z)\mathbf{A}^T\mathbf{x}\mathbf{x}^T\mathbf{A}\boldsymbol{\Sigma}(z) + \boldsymbol{\Sigma}(z)\right] \tag{35}$$

We make the ansatz that

$$\mathbf{C} = K\left(\mathbf{A}^T\mathbf{A}\right)^{-1} \tag{36}$$

where $K$ is an unknown constant. Substituting this guess into Eq. (34), we see that $\boldsymbol{\Sigma}(z)$ simplifies considerably:

$$\boldsymbol{\Sigma}(z) = \left(K^{-1} + \tfrac{z^2}{\sigma_\mathrm{x}^2}\right)^{-1}\left(\mathbf{A}^T\mathbf{A}\right)^{-1} \tag{37}$$

and as the data are whitened (assuming this is true at any contrast level, i.e. $\mathrm{E}_{\mathbf{x}|z}\left[\mathbf{x}\mathbf{x}^T\right] = c(z)\mathbf{I}$, with some $c(z)$), we indeed have

$$\mathrm{E}_{\mathbf{u}}\left[\mathbf{u}\mathbf{u}^T\right] \propto \left(\mathbf{A}^T\mathbf{A}\right)^{-1} \tag{38}$$

confirming our ansatz.

In principle, we could find $K$ by solving Eq. (35) (by substituting Eq. 36 to its l.h.s., and Eq. 37 to its r.h.s.), however, in practice, we cannot because we do not know $c(z)$ in $\mathrm{E}_{\mathbf{x}|z}\left[\mathbf{x}\mathbf{x}^T\right] = c(z)\mathbf{I}$. Instead, we set $K$ to ensure that the inputs, $\mathbf{A}^T\mathbf{x}$, have the right covariance (note that it is only possible to match the covariance of $\mathbf{A}^T\mathbf{x}$, and not of $\mathbf{x}$ directly, because we are using an undercomplete basis). As the data is whitened, we expect

$$\mathrm{E}\left[\mathbf{A}^T\mathbf{x}\mathbf{x}^T\mathbf{A}\right] = \mathbf{A}^T\mathbf{A} \tag{39}$$

while the predictive distribution of the GSM results in

$$\mathrm{E}\left[\mathbf{A}^T\mathbf{x}\mathbf{x}^T\mathbf{A}\right] = \mathbf{A}^T\left(\mathrm{E}\left[z^2\right]\mathbf{A}\mathbf{C}\mathbf{A}^T + \sigma_\mathrm{x}^2\mathbf{I}\right)\mathbf{A} \tag{40}$$

Setting these expressions equal, substituting for $\mathbf{C}$ using our ansatz (Eq. 36), and using $\mathrm{E}\left[z^2\right] = 1$ gives

$$\mathbf{A}^T\mathbf{A} = \left(K + \sigma_\mathrm{x}^2\right)\mathbf{A}^T\mathbf{A} \tag{41}$$

yielding the solution

$$K = 1 - \sigma_\mathrm{x}^2 \tag{42}$$

(Note that while this derivation is valid for the complete and undercomplete case, a more complex analysis would be necessary for the overcomplete case.)

With these choices, the dynamics only depend on the probabilistic model through the product $\left(\mathbf{A}^T\mathbf{A}\right)^{-1}$. This product controls the frequency spectrum: if $\left(\mathbf{A}^T\mathbf{A}\right)^{-1}$ has a very broad eigenspectrum (e.g. multiple orders of magnitude), then the system will sample at different rates along different directions. This is not desirable: we want sampling to take place as fast as possible in every direction, not to be fast in some directions, and slow in others. If we were able to set $\mathbf{M}$ to $\left(\mathbf{A}^T\mathbf{A}\right)^{-1}$, then we would indeed sample at the same rate in every direction [37], no matter how broad the spectrum of $\left(\mathbf{A}^T\mathbf{A}\right)^{-1}$ (see "Deriving the 1D approximate model", below). However, to ensure that Dale's law is obeyed, we need the elements of $\mathbf{M}$ to be non-negative, so we set

$$\mathbf{B} = \mathbf{I} \tag{43}$$

and

$$M_{ij} = \max\left(0, \left(\mathbf{A}^T\mathbf{A}\right)^{-1}_{ij}\right) \tag{44}$$

For the dynamics to be correct, we need this matrix to be positive definite. While this is not guaranteed, we found that in practice the matrix turns out to satisfy this constraint. As $\mathbf{M}$ is close to, but not exactly, $\left(\mathbf{A}^T\mathbf{A}\right)^{-1}$, the eigenspectrum of $\mathbf{A}^T\mathbf{A}$ will have some effect on our sampler. In practice, our eigenvalues range over a factor of 5 without weakening our results. Again, this is valid for the undercomplete and complete cases, and a more complex analysis would be necessary for the overcomplete case.

Next, we consider the observation noise level, $\sigma_x$, which describes the noise-to-signal ratio for neurons in the visual cortex. In particular, we take the input to be $\mathbf{A}^T\mathbf{x}$. This input is made up of two components, signal from the mean of $P(\mathbf{A}^T\mathbf{x}|\mathbf{u}, z)$, and noise from its covariance, (given by transforming Eq. (3)). The covariance of this input (Eq. 40) also breaks up into signal, $(1-\sigma_x^2)\mathbf{A}^T\mathbf{A}$, and noise, $\sigma_x^2\mathbf{A}^T\mathbf{A}$, terms, giving the signal to noise ratio as $\sqrt{\sigma_x^2/(1-\sigma_x^2)} \approx \sigma_x$. To obtain a value for $\sigma_x$ we perform a simple estimation. We take a V1 simple cell that integrates $N$ inputs from retinal ganglion cells (RGCs) (indirectly, via the LGN), each firing a Poisson spike train of average rate $r$, with a temporal integration window of $\Delta t$. In this case, the c.v. (which corresponds to $\sigma_x$) is

$$\sigma_x = \frac{\text{s.d.}}{\text{mean}} = \frac{\sqrt{Nr\Delta t}}{Nr\Delta t} = \frac{1}{\sqrt{Nr\Delta t}} \quad (45)$$

Based on the literature, we set the values of the relevant constants as

$$r \sim 1 \text{ s}^{-1} \text{ [72]}, \quad (46)$$
$$\Delta t \sim 10 \text{ to } 100 \text{ ms [73]}, \quad (47)$$
$$N \sim 100 \text{ to } 1000. \quad (48)$$

To obtain this range for $N$, we note that there are around 1000 RGCs in the stimulated region in [30]. (This can be computed knowing the dependency of RGC density on eccentricity [74], and that the stimulus has s.d. 0.5 degrees, so the total area is around 1 degree$^2$, and is 3 to 5 degrees from the fovea, and then discounting, to account for the fact that not all of these cells will be connected [75]). Thus, we obtain the interval

$$\sigma_x = \frac{1}{\sqrt{1}} \text{ to } \frac{1}{\sqrt{100}} \quad (49)$$

of which we use the geometric mean:

$$\sigma_x = \frac{1}{\sqrt{10}} \quad (50)$$

To choose values for $\tau_L$, $\tau$ and $\sigma_\mathbf{v}^2$, we considered biological constraints. The external input to the inhibitory cells is governed entirely by $\tau$, suggesting that a biologically plausible value for $\tau$ is 10 ms [76]. The scale of the recurrent input terms are governed by the product $\frac{1}{\tau}\mathbf{M}^{-1}$, suggesting that, to ensure the recurrent input has a biologically plausible timescale of 10 ms, we should set $\mathbf{M}^{-1}$ to be O(1) (see Eq. (44)).

Finally, we estimated $\tau_L$, or equivalently the amount of noise per unit time, by comparing the rate at which membrane potential variance increases in our equations, $2\sigma^2/\tau_L$, to the rate of increase given by stochastic vesicle release, the primary source of 'noise' in cortical circuits. If a neuron is connected to $s$ presynaptic neurons, firing with average rate $r$, and the variance of a unitary EPSP is $v$, then stochastic vesicle release introduces variance at the rate $srv$. Setting $srv = 2\sigma^2/\tau_L$ allows us to find the Langevin timescale

$$\tau_L = \frac{2\sigma^2}{srv} \qquad (51)$$

However, estimating $\tau_L$ is difficult, because there are huge uncertainties in $\sigma$, $s$, $r$ and $v$. We therefore wrote our uncertainty about each parameter as a log-normal distribution, $P(\log x) = \mathcal{N}\left(\log x; \mu_x, \sigma_x^2\right)$ where $x$ is one of $\sigma$, $s$, $r$, or $v$, and computed the induced distribution on $\tau_L$. To specify the distributions, we wrote a range, from $x_l$ to $x_h$, that, we believed contained around 95% of the probability mass, taking the boundaries of the range to be two standard-deviations from the mean in the log-domain, $\log x_l = \mu_x - 2\sigma_x$ and $\log x_h = \mu_x + 2\sigma_x$.

To estimate the required ranges, we took values from the neuroscience literature. First, estimates of firing rates vary widely, from around 0.5 Hz [77] to around 10 Hz [78]. Second, the number of synapses per cell is usually taken to be around 10000. However, it is likely that there are multiple synapses per connection [79], so there could be anywhere from 1000 to 10000 input cells for a single downstream neuron. Third, the average variance per spike is relatively easy to measure, data from Song *et al.* [80] put the value at 0.076 mV$^2$. As other measurements seem roughly consistent [81], we use a relatively narrow range for $v$, from 0.05 mV$^2$ to 0.1 mV$^2$. Finally, the scaling factor, $\sigma$, could plausibly range from 2.5 mV to 7.5 mV, giving a full (2 standard deviations, and both sides of the mean) range of membrane potential fluctuations of 10 mV to 30 mV [82].

These ranges give a central estimate of $\tau_L = 150$ ms, which we used in our simulations. In agreement with this back-of-the-envelope calculation, we find that our sampler's dynamics match neural dynamics when $\tau_L$ lies in a broad range, from around 60 ms to around 400 ms (see S1 Figure). While $\tau_L$ appears relatively large in comparison with typical neural timescales, which are often around 10 ms, it should be remembered that $\tau_L$ parameterises the amount of noise injected into the network at every time step, and as such, does not therefore have any necessary link to other neural time constants.

### Altering the model so that $u_i$ and $v_i$ are always positive

One might worry that it is possible for $u_i$ (or $v_i$) to go negative, meaning that they have their influence on downstream neurons will have the wrong sign. However, it is straightforward to offset $\mathbf{u}$ (and hence $\mathbf{v}$, through Eq. (12)), so that they rarely, if ever become negative. Moreover, if we introduce the offset as

$$P(\mathbf{u}) = \mathcal{N}(\mathbf{u}; \mathbf{b}, \mathbf{C}) \qquad (52)$$
$$P(\mathbf{x}|\mathbf{u}, z) = \mathcal{N}(\mathbf{x}; \mathbf{A}(\mathbf{u} - \mathbf{b}), \mathbf{C}) \qquad (53)$$

then this leaves the data distribution $P(\mathbf{x})$, and hence the dynamics intact.

## Deriving the 1D approximate model

$$\dot{\mathbf{u}} = \frac{1}{\tau}\mathbf{M}(\mathbf{u} - \mathbf{v}) \tag{54}$$

$$\dot{\mathbf{v}} = \frac{1}{\tau}\mathbf{M}(\mathbf{u} - \mathbf{v}) - \frac{z}{\sigma_x^2}\mathbf{A}^T(\mathbf{x} - z\mathbf{A}\mathbf{u}) - \mathbf{C}\mathbf{u} \tag{55}$$

Differentiating again yields

$$\ddot{\mathbf{u}} = \frac{1}{\tau}\mathbf{M}(\dot{\mathbf{u}} - \dot{\mathbf{v}}) \tag{56}$$

substituting for $\dot{\mathbf{u}}$ and $\dot{\mathbf{v}}$, and collecting the terms that depend on $\mathbf{u}$, we obtain

$$\ddot{\mathbf{u}} = -\frac{1}{\tau^2}\mathbf{M}\left(\frac{z^2}{\sigma_x^2}\mathbf{A}^T\mathbf{A} - \mathbf{C}^{-1}\right)(\mathbf{u} - \bar{\mathbf{u}}) \tag{57}$$

where $\bar{\mathbf{u}}$ is the posterior mean of $\mathbf{u}$ with fixed $z$ (see Eq. 33 37 and 42)

$$\bar{\mathbf{u}} = \frac{z}{\sigma_x^2}\left(\frac{z^2}{\sigma_x^2} + \frac{1}{1-\sigma_x^2}\right)(\mathbf{A}^T\mathbf{A})^{-1}\mathbf{A}^T\mathbf{x} \tag{58}$$

substituting $\mathbf{M} = (\mathbf{A}^T\mathbf{A})^{-1}$ (i.e. the ideal value for $\mathbf{M}$), and $\mathbf{C} = (1 - \sigma_x^2)(\mathbf{A}^T\mathbf{A})^{-1}$ (Eq. (36)), gives

$$\ddot{\mathbf{u}} = -\frac{1}{\tau^2}\left(\frac{z^2}{\sigma_x^2} + \frac{1}{1-\sigma_x^2}\right)(\mathbf{u} - \bar{\mathbf{u}}) \tag{59}$$

Thus, for fixed $z$, each component of $\mathbf{u}$ evolves independently.

## Simulation Protocol

We simulated stimulus onset by first running the sampler until it reached equilibrium with no stimulus, then turning on the stimulus. To represent no stimulus we sampled $\mathbf{x}$ from $P(\mathbf{x}|z=0)$, and to represent stimulus, we sampled $\mathbf{x}$ from $P(\mathbf{x}|z=z_{\text{gen}})$, where $z_{\text{gen}} \in \{0.5, 1, 2\}$.

## Computing LFPs and firing rates

To make contact with experimental data, we also computed local field potentials (LFPs), and firing rates. There are many methods for computing LFPs, we chose the simplest, averaging the membrane potentials across neurons, as it gave similar results to the other methods, without tuneable parameters. To compute firing rates, we used a rectified linear function of the membrane potential:

$$f_i(t) = \begin{cases} u_i(t) & \text{if } u_i(t) > 0 \\ 0 & \text{otherwise} \end{cases} \qquad (60)$$

# Acknowledgements

We thank G. Orbán for useful discussions and suggestions. This work was supported by the Wellcome Trust (ML), the Gatsby Charitable Foundation (LA), and the European Union Seventh Framework Programme (FP7/2007-2013) under grant agreement no. 269921 (BrainScaleS) (ML).

## Supplementary Figure

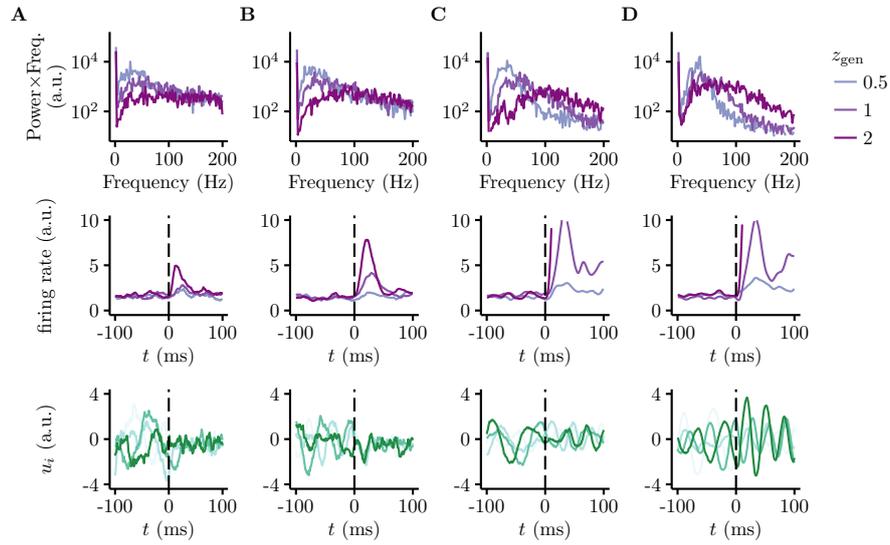

**Figure 8.** Our main results are robust to a range of $\rho$ or equivalently $\tau_L$. The top row is a power spectrum, the middle row displays the firing rate transient at stimulus onset, and the bottom row displays the membrane potential at stimulus onset for multiple trials and one neuron. The different lines in the first two rows correspond to different values of $z_{\text{gen}}$. In the bottom row, different lines correspond to different trials. **A.** For $\tau_L = 30$ ms, transients are small or non-existent, and no clear trends are present in the peak frequency. **B-C.** For $\tau_L = 60$ ms (**B**), and $\tau_L = 400$ ms (**C**) the results are similar to those in the main text. **D.** For $\tau_L = 1000$ ms, the results are quite different to those in the main text. In particular, the transient at stimulus onset lasts a long time, certainly longer than the observed value of around 50 ms.